\documentclass[floatfix,aps,pre,showpacs,showkeys,groupedaddress,twocolumn]{revtex4}
\usepackage{graphicx}
\usepackage{amsmath}

\begin{document}

\title{$T \rightarrow 0$ mean-field population dynamics approach for the random $3$-satisfiability problem\footnote{This paper was published in Physical Review E  {\bf 77} (2008) 066102.}}
\author{Haijun Zhou}
\affiliation{Institute of Theoretical Physics,Chinese Academy of Sciences, 
  Beijing 100080, China}

\begin{abstract}
  During the past decade, phase-transition phenomena in the random $3$-satisfiability ($3$-SAT)
  problem has been intensively studied by statistical physics methods. In this work,
  we study the random $3$-SAT problem by the mean-field first-step
  replica-symmetry-broken cavity theory at the limit of temperature $T\rightarrow 
  0$. The reweighting parameter $y$ of the cavity theory is allowed to approach 
  infinity together with the inverse temperature $\beta$  with
  fixed ratio $r=y / \beta$. Focusing on the the system's space of satisfiable configurations,
  we carry out extensive population dynamics simulations using the
  technique of importance sampling and we obtain the entropy density $s(r)$ and
  complexity $\Sigma(r)$ of zero-energy clusters at
  different $r$ values. We demonstrate that the population
  dynamics may reach different fixed points with different types of initial conditions.
  By knowing the trends of $s(r)$ and $\Sigma(r)$ with $r$, we can judge whether
  a certain type of initial condition is appropriate at a given $r$ value.
  This work complements and confirms the results of several other very recent theoretical studies.
\end{abstract}


\maketitle

\section{Introduction}

Critical behaviors in the random $3$-satisfiability ($3$-SAT)
problem were first reported by
Kirkpatrick and Selman in 1994 \cite{Kirkpatrick-Selman-1994}. Since then,
physicists working in the field of spin glasses have done a lot of work
on this important model system in theoretical computer science
\cite{Monasson-Zecchina-1996,Monasson-Zecchina-1997}. 
Mean-field calculations were done to understand the nature of the satisfiability
(SAT-UNSAT) transition 
\cite{Monasson-Zecchina-1996,Monasson-Zecchina-1997,Monasson-etal-1999,Zhou-2005b},
to locate the SAT-UNSAT transition point
\cite{Mezard-etal-2002,Mezard-Zecchina-2002,Mertens-etal-2006}, and to
analyze the performances of various algorithms
\cite{Cocco-Monasson-2001}. 
Based on the first-step replica-symmetry-broken (1RSB) mean-field cavity theory 
of spin glasses \cite{Mezard-Parisi-2001}, M{\'{e}}zard, Parisi, and Zecchina 
created a powerful message-passing algorithm, namely survey propagation (SP), to
find satisfiable solutions to random $3$-SAT formulas \cite{Mezard-etal-2002}. 
The physical picture underlying the SP algorithm is that, when the density of
constraints $\alpha$ of the system 
is close to the satisfiability threshold $\alpha_{\rm s}$,
the solution space of a random $3$-SAT formula divides into many
well-separated clusters. 
M{\'{e}}zard and co-workers also predicted
that the SAT-UNSAT transition for the random $3$-SAT problem occurs at 
$\alpha_{\rm s} = 4.2667$ \cite{Mezard-Zecchina-2002,Mertens-etal-2006}.
This threshold value lies within the
rigorously known lower-bound $3.52$ \cite{Hajiaghayi-Sorkin-2003}
and upper-bound $4.506$ \cite{Dubois-etal-2000} for random $3$-SAT, and
the mean-field cavity SP
solution is locally stable \cite{Montanari-etal-2004,Mertens-etal-2006,Zhou-etal-2007}.
The predicted SAT-UNSAT transition point of $\alpha_{\rm s}=4.2667$ is therefore 
conjectured to be exact.

The message-passing SP algorithm corresponds to the
temperature $T=0$ (i.e., $\beta = 1/ T = +\infty$) limit of the 1RSB mean-field cavity theory
of finite-connectivity spin glasses \cite{Mezard-Parisi-2001,Mezard-Parisi-2003}.
This 1RSB cavity theory has an adjustable reweighting parameter $y$. 
In Refs.~\cite{Mezard-etal-2002,Mezard-Zecchina-2002,Mertens-etal-2006}, first the inverse
temperature $\beta$ is set to infinity, and then $y$ is set to infinity. This means that
the ratio $\lim_{T\rightarrow 0} y / \beta$ is equal to zero. On the other hand,
it is now recognized that, to correctly characterize the
equilibrium properties (as represented by the free-energy Gibbs measure) of
a spin glass system, the reweighting parameter $y$ is required to take an
appropriate value that is dependent on $\beta$. 
For a spin glass system with many-body interactions, there
may exist a temperature range $T_{\rm d} \geq  T \geq  T_{\rm c}$
within which the optimal value of the
reweighting parameter  $y$ is equal to $\beta$
\cite{Kirkpatrick-Thirumalai-PRB-1987,Monasson-1995,Zhou-Li-CTP-2008}.
In the literature on structural glasses \cite{Kirkpatrick-Thirumalai-PRB-1987},
$T_{\rm d}$ and $T_{\rm c}$ are referred to as the
dynamical and static transition temperature of the system, respectively.
For the random $3$-SAT problem with density of constraints $\alpha$,
if the corresponding static transition temperature
is located at $T_{\rm c}(\alpha)=0$, then the reweighting parameter $y$ and the
inverse temperature $\beta$ should approach infinity with the same rate. 
In the present work, we investigate how the mean-field predictions on the
ground-state properties of the random $3$-SAT problem depend on the
ratio $y/ \beta$. 
We generalize the cavity treatment of 
Refs.~\cite{Mezard-etal-2002,Mezard-Zecchina-2002,Mertens-etal-2006} and study 
the statistical mechanics properties of the random $3$-SAT problem 
in the limit  $\beta \rightarrow +\infty$ and $y \rightarrow +\infty$, with 
fixed ratio \cite{Mezard-etal-2005}
\begin{equation}
  \label{eq:Ratio}
  r \equiv \frac{y}{\beta} \ .
\end{equation}
Population dynamics simulations were performed based on
a set of mean-field 1RSB cavity equations, and for each value of
$\alpha$,  the entropy density $s(r)$ and complexity $\Sigma(r)$
of the system as a function of
the ratio $r$ are estimated. 
The entropy density $s(r)$ is a measure of the number of ground-energy configurations
within one cluster of the configuration space, while the complexity $\Sigma(r)$ is a measure
of the total number of such ground-energy clusters.

As the population dynamics simulations of this work were running,
we noticed that questions closely related to
the issue we discuss here were investigated earlier in
Ref.~\cite{Mezard-etal-2005} in the context of the random $3$-coloring problem and
more recently in Refs.~\cite{Krzakala-etal-PNAS-2007,Zdeborova-Krzakala-PRE-2007}
for random $q$-coloring and
random $K$-SAT. While the main focus of Ref.~\cite{Krzakala-etal-PNAS-2007} was on
the limiting case of $r=1$, at which the numerical complexity of
the mean-field theory can be reduced to some extent, detailed discussions on general
values of $0 \leq r \leq 1$ were presented in
Refs.~\cite{Zdeborova-Krzakala-PRE-2007,Montanari-etal-2008}. 
The present paper confirms the physical picture given by Krzakala, Montanari, and co-workers
\cite{Krzakala-etal-PNAS-2007,Zdeborova-Krzakala-PRE-2007,Montanari-etal-2008} on the
solution space structure of random $3$-SAT; it is complementary to these
theoretical studies in three important ways. First, we introduce
a different scheme of population dynamics with importance sampling (this scheme can
be readily extended to finite temperatures); the numerical results obtained from this
scheme are in agreement with those reported in Ref.~\cite{Montanari-etal-2008}.
Second, we demonstrate that the population dynamics may reach different
fixed points from different initial conditions. Third, we find that different
initial conditions will lead to the {\em same} prediction on the properties of the
dominating solution clusters of random $3$-SAT. This last point is rather
interesting and needs to be further studied.

The main results of this paper are summarized here.
When using the $F$-type initial condition as
described in Sec.~\ref{sec:populationdynamics}, the population dynamics demonstrates that
(i) at $\alpha = \alpha_s = 4.2667$, 
$\Sigma(r)$ decreases monotonically with $r$ according to
$\Sigma(r)= - 0.020 r^2$ and $s(r)$ increases monotonically with $r$; (ii)
at $\alpha = 4.2$, the complexity changes with $r$ following
$\Sigma(r) = 0.0059-0.023 r^2$ and $s(r)$ still increases monotonically
with $r$; (iii) at $\alpha = 4.0$, both $\Sigma(r)$ and $s(r)$ have a discontinuity at
$r=0$. When using the $U$-type initial condition of Sec.~\ref{sec:populationdynamics}, we
find that both $\Sigma(r)$ and $s(r)$ are not monotonic functions of $r$.
At the value of $r=1$, the complexity $\Sigma(1)$ and entropy density
$s(1)$ as a function of the constraint density $\alpha$ are also calculated
by population dynamics simulations with both the $F$-type and the $U$-type initial
condition. The numerical data are consistent with the conclusion of
Ref.~\cite{Krzakala-etal-PNAS-2007} that, for $\alpha < 3.87$ the solution space of the
random $3$-SAT problem forms a single cluster, while for $3.87 \leq \alpha < \alpha_{\rm s}$
the solution space, although being nonergodic,
is dominated by only a few (of order unity) solution clusters.

The paper is organized asfollows.
Section \ref{sec:Method} describes the mean-field cavity approach and the protocol of
population dynamics simulations. The simulation results
are reported and analyzed in Sec.~\ref{sec:RD}. We conclude our
work in Sec.~\ref{sec:CO} and discuss possible future extensions.

\section{Method}
\label{sec:Method}

\subsection{The factor-graph representation of the random $3$-SAT problem}

\begin{figure}
	\includegraphics[width=1.0\linewidth]{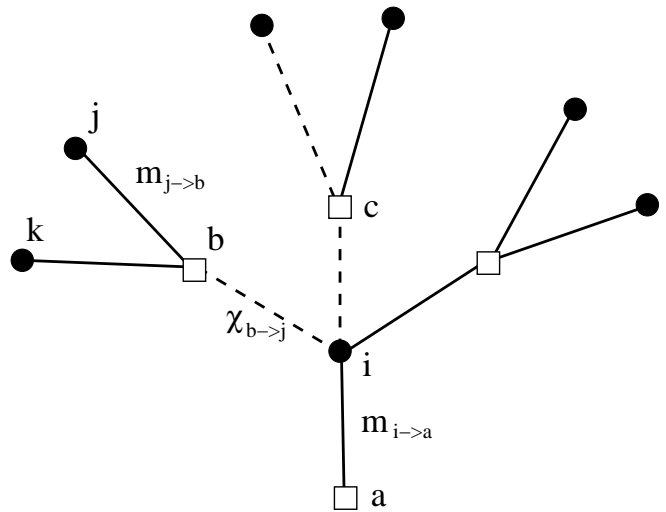}
  \caption{\label{fig:factor}
    The factor-graph representation 
    \cite{Kschischang-etal-2001,Mezard-etal-2002,Mezard-Zecchina-2002}
    for a random $3$-SAT formula. $m_{j\rightarrow b}$, $m_{i \rightarrow a}$ and $\chi_{b \rightarrow i}$ are 
    messages on the edges of the factor-graph   (explained in the main 
    text).
  }
\end{figure}

A $3$-SAT formula contains $N$ Boolean variables and $M$ constraints, each of
which involves $K=3$ variables.
The degree of
constrainedness of a random $3$-SAT formula is characterized by the
constraint density $\alpha \equiv M / N$.
A $3$-SAT formula can be represented by a factor graph 
${\cal G}$ 
(see Fig.~\ref{fig:factor}) of $N$ variable nodes (circles $i,j,k,\ldots$) and
$M$ function nodes (squares $a,b,c,\ldots$)
\cite{Kschischang-etal-2001,Mezard-etal-2002,Mezard-Zecchina-2002}. 
Each function node $a$ corresponds to a constraint; it is connected to
$K$ ($=3$) variable nodes $i \in \partial a$ 
(where $\partial a$ 
denotes the set of nearest neighbors of node $a$). 
Associated with each function 
node $a$  is an energy $E_a \in \{0, 2 \}$ of the form
\begin{equation}
  \label{eq:Ea} 
  E_a = 2 \prod\limits_{i\in \partial a} \frac{1-J_a^i \sigma_i}{2} \ .
\end{equation}
In Eq.~(\ref{eq:Ea}), $\sigma_i = \pm 1$ is the spin value of variable node $i$;
$J_a^i = \pm 1$ is the coupling between node $i$ and node $a$. In the factor
graph, the edge $(i,a)$ is a solid line if $J_a^i=1$ and it is a
dashed line if $J_a^i=-1$.  
For a given $3$-SAT formula, the factor graph
(with all its coupling constants)
is fixed, while the spin configuration ${\bf \sigma} \equiv \{ \sigma_1, \sigma_2, \ldots, \sigma_N \}$
can change. The total energy $E({\bf \sigma})$ of a given spin configuration
is
\begin{equation}
  \label{eq:Hamiltonian}
  E({\bf \sigma} ) = \sum\limits_{a \in {\cal G}} E_a \ .
\end{equation}

A variable node $i$ of the factor graph ${\cal G}$ 
is connected to $k_i$ function nodes $a \in \partial i$. 
the vertex degree $k_i$ may be different for different variable nodes.
For a random
$3$-SAT formula with $N \gg 1$, the distribution of $k_i$ is governed by
the
Poisson distribution of mean $3 \alpha$, i.e., ${\rm Prob}(k_i=k) = f_{3 \alpha}(k) \equiv
(3 \alpha)^k e^{-3 \alpha} / k!$. One can also define the $``$cavity degree'' $k_{i\rightarrow a}$ 
of a variable node $i$ with respect to an edge $(a,i)$ as 
$k_{i\rightarrow a} \equiv | \partial i \backslash a |$. $k_{i \rightarrow a}$ is the number of nearest neighbors of node $i$ when
edge $(a,i)$ is not considered. Obviously, $k_{i\rightarrow a} = k_i - 1$. 
A useful property of random graphs is that the distribution of
$k_{i\rightarrow a}$ is also governed by the Poisson distribution of mean $3 \alpha$.
We will use this property in the mean-field population dynamics simulations
as described in Sec.~\ref{sec:populationdynamics}.

\subsection{The cavity equations at a general low temperature $T$}

At a sufficiently low temperature $T$, ergodicity of the whole
configurational space $\Lambda$ of the
model Eq.~(\ref{eq:Hamiltonian}) breaks down. It is then
assumed in mean-field theories 
\cite{Mezard-Parisi-2001,Mezard-etal-2002,Mezard-Zecchina-2002} that
$\Lambda$ is split into an exponential number of ergodic
subspaces. Each of these subspaces $\Lambda_\alpha$ corresponds to a
macroscopic state (macrostate $\alpha$) of the system at temperature $T$.
Based on the cavity approach of spin glasses
\cite{Mezard-Parisi-2001,Mezard-etal-1987}, the mean
grand free-energy density of the random $3$-SAT problem can be
derived. As the derivation details are well documented in the
literature \cite{Mezard-Parisi-2001,Mezard-Zecchina-2002}
(see also Refs.~\cite{Zhou-2007b,Zhou-Li-CTP-2008}),
we shall directly list the final expressions and give only
brief explanations.

At the 1RSB level of approximation, the total grand
free energy of the random $K$-SAT system is
\begin{equation}
  \label{eq:GrandFreeEnergy}
  G_{\cal G}(\beta; y)= \sum\limits_{i\in {\cal G}} \Delta G_i - (K-1) \sum\limits_{a \in {\cal G}} \Delta G_a \ ,
\end{equation}
where $\Delta G_i$ and $\Delta G_a$ are, respectively,
the grand free-energy increase caused by adding variable node $i$ and function node 
$a$, with
\begin{equation}
  \label{eq:DeltaGi}
  \Delta G_i = -\frac{1}{y} \log\biggl[ 
  \prod\limits_{b \in \partial i} \bigl(\int {\rm d} {\chi}_{b\rightarrow i} \hat{P}_{b\rightarrow i}({\chi}_{b\rightarrow i}) \bigr)
  \exp(-y \Delta F_i ) 
  \biggr]
\end{equation}
and
\begin{equation}
  \label{eq:DeltaGa}
  \Delta G_a =
 - \frac{1}{y} \log\biggl[
  \prod\limits_{j\in \partial a} \bigl( \int {\rm d} m_{j\rightarrow a} P_{j\rightarrow a}(m_{j\rightarrow a}) \bigr)
  \exp(-y \Delta F_{a} ) 
  \biggr] \ .
\end{equation}
In Eqs.~(\ref{eq:DeltaGi}) and (\ref{eq:DeltaGa}), $m_{i\rightarrow a}$ (the cavity magnetization)
is the mean magnetization of vertex $i$ within one macrostate $\alpha$ when the
edge $(a,i)$ is discarded, and $P_{i\rightarrow a}(m_{i\rightarrow a})$ is the distribution of
this cavity magnetization among all the macrostates of the system. Similarly,
$\chi_{b\rightarrow i} \equiv \prod_{j\in \partial b\backslash i}[(1-J_b^j m_{j\rightarrow b})/2]$ is the directed message from function
node $b$ to variable node $i$ in one macrostate, and $\hat{P}_{b\rightarrow i}({\chi}_{b\rightarrow i})$ is the
distribution of this message among all the macrostates. $\Delta F_i$ and $\Delta F_a$ are,
respectively, the free-energy increase of macrostate $\alpha$ due to the addition
of variable node $i$ and function node $a$, with
\begin{eqnarray}
  \Delta F_i & = &- \frac{1}{\beta} \log\biggl[{\prod\limits_{b\in \partial i}^{(-)}} \bigl[ 1- (1-e^{-2 \beta}) {\chi}_{b\rightarrow i} \bigr] \nonumber \\
& & 
  + {\prod\limits_{b\in \partial i}^{(+)}} \bigl[ 1- (1-e^{-2 \beta}) {\chi}_{b\rightarrow i} \bigr] \biggr] \ ,
  \label{eq:DeltaFi} \\
  \Delta F_a & = &- \frac{1}{\beta} \log\biggl[1- (1-e^{-2 \beta}) \prod\limits_{j\in \partial a} \frac{(1-J_a^j m_{j\rightarrow a})}{2}
  \biggr] \ .
  \label{eq:DeltaFa}
\end{eqnarray}
In Eq.~(\ref{eq:DeltaFi}), the $\prod\limits^{(-)}$ and $\prod\limits^{(+)}$ indicate that
the multiplication is restricted to the neighbors $b$ of $i$ for 
which $J_b^i = -1$ and $J_b^i=+1$, respectively.  

On each edge $(a,i)$ of factor graph ${\cal G}$, the
probability distributions $P_{i\rightarrow a}$ and $\hat{P}_{a\rightarrow i}$ are 
required to satisfy the variational condition that
\begin{equation}
  \label{eq:VariationalCondition}
  \frac{ \delta G_{{\cal G}}(\beta; y) }{\delta P_{i\rightarrow a} } =
  \frac{ \delta G_{{\cal G}}(\beta; y) }{\delta \hat{P}_{a\rightarrow i} } \equiv 0 \ .
\end{equation}
This variational condition is satisfied 
by the following two self-consistent equations on each directed edge $a\rightarrow i$ and
$i\rightarrow a$:
\begin{widetext}
\begin{equation}
  \label{eq:hatPai}
  \hat{P}_{a\rightarrow i} ({\chi}_{a\rightarrow i}) = \prod\limits_{j\in \partial a\backslash i} \bigl[ \int {\rm d} m_{j\rightarrow a}
  P_{j\rightarrow a}(m_{j\rightarrow a}) \bigr] \delta \biggl( \chi_{a\rightarrow i} - 
  \prod\limits_{j\in \partial a\backslash i} \frac{(1-J_a^j m_{j\rightarrow a})}{2} \biggr) 
\end{equation}
and
\begin{equation}
  \label{eq:Pia}
  P_{i\rightarrow a}(m_{i\rightarrow a}) =
 \frac{ \prod\limits_{b \in \partial i \backslash a} \bigl[ \int {\rm d} \chi_{b\rightarrow i} \hat{P}_{b\rightarrow i}(\chi_{b\rightarrow i}) \bigr]
    e^{-y \Delta F_{i\rightarrow a}} \delta\bigl( m_{i\rightarrow a}- M(\{ \chi_{b\rightarrow i}: b\in \partial i\backslash a \} \bigr) }{
    \prod\limits_{b \in \partial i \backslash a} \bigl[ \int {\rm d} \chi_{b\rightarrow i} \hat{P}_{b\rightarrow i}(\chi_{b\rightarrow i}) \bigr]
    e^{-y \Delta F_{i\rightarrow a}} } \ ,
\end{equation}
with $M(\{{\chi}_{b\rightarrow i}: b\in \partial i\backslash a \})$ being the shorthand notation for
\begin{equation}
  \label{eq:M}
  M(\{ \chi_{b\rightarrow i} : b\in \partial i\backslash a \}) \equiv 
  \frac{ {\prod\limits_{b\in \partial i\backslash a}^{(-)}}\bigl[1-(1-e^{-2 \beta}) \chi_{b\rightarrow i} \bigr]-{\prod\limits_{b\in \partial i\backslash a}^{(+)}}
    \bigl[1-(1-e^{-2 \beta}) \chi_{b\rightarrow i} \bigr]}
  {
    {\prod\limits_{b\in \partial i\backslash a}^{(-)}}
    \bigl[1-(1-e^{-2 \beta}) \chi_{b\rightarrow i} \bigr]+{\prod\limits_{b\in \partial i\backslash a}^{(+)}}
    \bigl[1-(1-e^{-2 \beta}) \chi_{b\rightarrow i} \bigr]} \ .
\end{equation}
\end{widetext}
The free-energy increase $\Delta F_{i\rightarrow a}$ 
in Eq.~(\ref{eq:Pia}) is calculated by Eq.~(\ref{eq:DeltaFi}) but with
$b\in \partial i$ being replaced by $b\in \partial i\backslash a$ [i.e., discarding the effect of edge $(i,a)$].

\subsection{The $T\rightarrow 0$ limit and population dynamics simulations}
\label{sec:populationdynamics}

Let us now consider the zero-temperature limit 
(i.e., $\beta \rightarrow +\infty$) of the cavity equations of the
preceding subsection. We focus on the SAT phase of the random $3$-SAT
problem and assume the 
Hamiltonian Eq.~(\ref{eq:Hamiltonian}) has at least one zero-energy ground state.
In the SAT phase at the $\beta\rightarrow +\infty$ limit, the free energy of each macrostate is
completely contributed by entropy. 

For the benefit of later discussions, 
let us introduce two further shorthand notations $Z_i$ and $Z_a$,
\begin{eqnarray}
  Z_i & \equiv &  {\prod\limits_{b\in \partial i}^{(-)}} \bigl(1- {\chi}_{b\rightarrow i} \bigr)
  + {\prod\limits_{b\in \partial i}^{(+)}} \bigl(1- {\chi}_{b\rightarrow i} \bigr)  \ , 
  \label{eq:Zi} \\
  Z_a & \equiv & 1- \prod\limits_{j\in \partial a} \frac{(1-J_a^j m_{j\rightarrow a})}{2} \ .
  \label{eq:Za}
\end{eqnarray}
Then at $\beta \rightarrow \infty$ and fixed ratio $r$,
the grand free-energy increases $\Delta G_i$ and $\Delta G_a$ can be reexpressed
as
\begin{widetext}
\begin{eqnarray}
  y \Delta G_i &= &  - \log\Bigl[ \prod\limits_{b\in \partial i} \bigl(\int {\rm d} {\chi}_{b\rightarrow i} 
   \hat{P}_{b\rightarrow i}({\chi}_{b\rightarrow i}) \bigr) \Theta(Z_i) e^{ r \log(Z_i) }  \Bigr] \ ,
  \label{eq:yDeltaGi} \\
  y \Delta G_a &=& - \log\Bigl[\prod\limits_{i\in \partial a} \bigl(\int {\rm d} {m}_{i\rightarrow a} P_{i\rightarrow a}({m}_{i\rightarrow a}) \bigr) \Theta(Z_a) 
  e^{ r \log ( Z_a ) } \Bigr] \ .
  \label{eq:yDeltaGa}
\end{eqnarray}
In Eqs.~(\ref{eq:DeltaGi}) and (\ref{eq:DeltaGa}), $\Theta(x) = 1$ if $x >0$ and $\Theta(x)=0$ if
$x \leq 0$. 

In the thermodynamic limit of graph size $N\rightarrow \infty$ and $M\rightarrow \infty$ (with $\alpha$ being finite), 
the grand free-energy density $g(r)$ of the random $3$-SAT system is expressed
as
\begin{equation}
  \label{eq:meanGFEdensity}
  y g(r) =  y \overline{ \Delta G_i} - 2 \alpha  y \overline{ \Delta G_a} \ ,
\end{equation}  
where the overlines indicate averaging over all the possible local environments of
the involved variable node $i$ or function node $a$.
The complexity $\Sigma(r)$ and mean entropy density $s(r)$ of the system are related to
$g(\beta; y)$ by (see, e.g., Ref.~\cite{Zhou-Li-CTP-2008})
\begin{eqnarray}
  \Sigma(r) & = & - y g(r) + r \Bigl( \overline{ \langle \beta  \Delta F_i \rangle } -
  2 \alpha  \overline{ \langle \beta \Delta F_a \rangle } \Bigr)  \ , \label{eq:Complexity} \\
  s(r) & = & - \Bigl( \overline{ \langle \beta  \Delta F_i \rangle } -
  2 \alpha  \overline{ \langle \beta \Delta F_a \rangle } \Bigr) \ . \label{eq:EntropyDensity}
\end{eqnarray}
The mean free-energy increase of $\Delta F_i$ and $\Delta F_a$ as averaged over all the
macrostate of the system is
calculated through
\begin{eqnarray}
  \langle \beta \Delta F_i \rangle & = & -  \frac{\prod\limits_{b\in \partial i} \bigl[\int {\rm d} {\chi}_{b\rightarrow i} \hat{P}_{b\rightarrow i}({\chi}_{b\rightarrow i}) \bigr]
    \Theta(Z_i) e^{ r \log(Z_i)} \log( Z_i)  }{
    \prod\limits_{b\in \partial i} \bigl[\int {\rm d} {\chi}_{b\rightarrow i} \hat{P}_{b\rightarrow i}({\chi}_{b\rightarrow i}) \bigr] \Theta(Z_i) 
    e^{ r \log(Z_i)} } \ ,
  \label{eq:meanDeltaFi} \\
  \langle \beta \Delta F_a \rangle & = & - \frac{\prod\limits_{i\in \partial a} \bigl[\int {\rm d} {m}_{i\rightarrow a} {P}_{i\rightarrow a}({m}_{i\rightarrow a}) \bigr]
    \Theta(Z_a) e^{ r \log(Z_a)} \log( Z_a)  }{
    \prod\limits_{i\in \partial a} \bigl[\int {\rm d} {m}_{i\rightarrow a} {P}_{i\rightarrow a}({m}_{i\rightarrow a}) \bigr] \Theta(Z_a) 
    e^{ r \log(Z_a)} } \ .
  \label{eq:meanDeltaFa}
\end{eqnarray}
\end{widetext}

At a given value of constraint density $\alpha$, 
we use population dynamics \cite{Mezard-Parisi-2001} to calculate the
complexity $\Sigma(r)$ and entropy density $s(r)$ for the random $3$-SAT problem.
The iterative equation (\ref{eq:Pia}) for the
cavity magnetization distributions $P_{i\rightarrow a}(m_{i\rightarrow a})$ are 
implemented according to the following protocol of importance sampling.
\begin{description}
\item[(1)]. A total number of ${\cal N}$ sets are stored
  in the computer memory. 
  Each set, which represents a probability distribution
  $P_{i\rightarrow a}(m_{i\rightarrow a})$ of a cavity magnetization,
  contains ${\cal M}$ double-precision values $-1 \leq m_{i\rightarrow a} \leq 1$. 
  These ${\cal N}$ sets are independently initialized according to
  a certain type of initial condition (see below).
  
\item[(2)]. To perform a single update to the stored population of distributions, the
  follow steps occur:
  
  \begin{description}
  \item[(i)] A random integer $n$ is generated
    according to the Poisson distribution $f_{3 \alpha}(n)$.
    
  \item[(ii)] $ 2 n$ sets 
    (denoted by $P_{j_1\rightarrow  b_1}, P_{k_1\rightarrow b_1}, \ldots, P_{j_{n}\rightarrow b_n}, P_{k_{n}\rightarrow b_n}$) 
    are randomly chosen with replacement from the stored 
    ${\cal N}$ sets, and $3 n$ coupling constants $\{ J_{b_l}^{j_l}, J_{b_l}^{k_l}, 
    J_{b_l}^i \}$ are generated, each of which is independently 
    assigned a value $+1$ or $-1$ with probability one-half.
    
  \item[(iii)] $2 n$ cavity magnetizations 
    $(m_{j_1 \rightarrow b_1}, m_{k_1\rightarrow b_1})$, $\ldots$, $(m_{j_n \rightarrow b_n}, m_{k_n \rightarrow b_n})$ 
    are sampled uniformly from these $2 n$ sets, respectively, 
    and $n$ values $\chi_{b_l \rightarrow i}= [(1-J_{b_l}^{j_l} m_{j_l \rightarrow b_l})
    (1-J_{b_l}^{k_l} m_{k_l \rightarrow b_l} )/ 4$ are calculated.
    
  \item[(iv)] $Z_i = \prod_{b_l: J_{b_l}^i=-1} (1- {\chi}_{b_l\rightarrow i}) +\prod_{b_l: J_{b_l}^i=+1} (1- {\chi}_{b_l\rightarrow i})$ 
    is calculated and a new cavity magnetization 
    $m_{i\rightarrow a} = [\prod_{b_l: J_{b_l}^i=-1} (1- {\chi}_{b_l\rightarrow i}) -
    \prod_{b_l: J_{b_l}^i=+1} (1- {\chi}_{b_l\rightarrow i}) ]/Z_i$ is calculated.
    
  \item[(v)] This new $m_{i\rightarrow a}$ value is accepted with probability proportional to
    $\Theta(Z_i) \exp[r \log(Z_i)]$ by way of the Metropolis importance-sampling
    method \cite{Newman-Barkema-1999} and, if it is rejected, the old $m_{i\rightarrow a}$ value
    is retained.

  \item[(vi)] Repeat ({\bf iii})--({\bf v})
    a number of ${\cal M} \times {\cal L}$ times and generate a new set $P_{i\rightarrow a}$
    with ${\cal M}$ independent $m_{i\rightarrow a}$ values (sampled with
    interval ${\cal L}$). Obtain the value of $y \Delta G_i$ and $\langle \beta \Delta F_{i} \rangle$ 
    as expressed by  Eqs.~(\ref{eq:yDeltaGi}) and (\ref{eq:meanDeltaFi})
    using the sampled data of these ${\cal M} \times {\cal L}$ repeats.

  \item[(vii)] Replace a randomly chosen stored old set with the newly generated
    set $P_{i\rightarrow a}$.
  \end{description}

\item[(3)]. Repeat step (${\bf 2}$) three times to obtain three probability distributions
$P_{i \rightarrow a}(m_{i \rightarrow a})$, $P_{j\rightarrow a}(m_{j \rightarrow a})$,
 and $P_{k \rightarrow a}(m_{k \rightarrow a})$. A total number of ${\cal M} \times
{\cal L}$ triples $(m_{i\rightarrow a}, m_{j\rightarrow a}, m_{k\rightarrow a})$ are then sampled uniformly. From these
sampled data, $y \Delta G_a$ and $\langle \beta \Delta F_a \rangle$ as expressed by
Eqs.~(\ref{eq:yDeltaGa}) and (\ref{eq:meanDeltaFa}) are calculated.

\item[(4)]. Repeat steps (${\bf 2}$) and (${\bf 3}$) a number ${\cal T}_1  \times {\cal N}$ of 
times for the population dynamics to reach a steady 
state and another number ${\cal T}_2 \times {\cal N}$ of times to collect
values of $y \Delta G_i$, $y \Delta G_a$, $\langle \beta \Delta F_i \rangle$,
 and $\langle \beta \Delta F_a \rangle$. From these
collected values, the grand free-energy density $g(r)$, the complexity $\Sigma(r)$, and
the mean entropy density $s(r)$ are calculated according to Eqs.~(\ref{eq:meanGFEdensity}),
(\ref{eq:Complexity}), and (\ref{eq:EntropyDensity}), respectively. The standard deviations of
the numerical results are estimated by the bootstrap method \cite{Efron-SIAM-1979}.
\end{description}

The above-mentioned population dynamics procedure is quite time-consuming. The total
simulation time is roughly proportional to ${\cal N} {\cal M} {\cal L} ({\cal T}_1 + {\cal T}_2 )$. 
We have used different sets of parameter values to reach a balance between high numerical precision and
computation time. The data reported in the next section are obtained with the following set of
parameters: ${\cal N}=1000$, ${\cal M}=2000$, ${\cal L}=50$, ${\cal T}_1 = 500$, 
and ${\cal T}_2 = 1500$ (with the exception that, in Fig.~\ref{fig:R1}, the simulation results
at $\alpha=3.875$ and $3.9$, which are close to the
ergodicity transition point of the random $3$-SAT system, are  obtained
with ${\cal N}=2016$ and ${\cal L}=100$). At each pair of values $(\alpha, r)$, 
this set of parameters leads to satisfactory numerical precision
with a tolerable simulation time of about ten days (through a present-day personal 
computer). If we use ${\cal N}=2000$ and ${\cal M}=4000$ in the simulation, the mean values of
the calculated thermodynamic densities will not change much, while their standard deviations
can be reduced to about half the level of those reported in the next section.

It is recognized by test runs that the results of the population dynamics can have a strong
dependence on initial condition. The set of self-consistent equations (\ref{eq:hatPai}) and
(\ref{eq:Pia}) for the random $3$-SAT problem may have more than one stable fixed point. To
investigate this initial condition dependence, we use the following
two major types of initial conditions
to produce numerical data of the next section.
\begin{description}
\item[{\it F-type}.] The cavity magnetization distribution $P_{i \rightarrow a} (m_{i\rightarrow a})$ at the beginning of the
  population dynamics is set to be $P_{i\rightarrow a} (m_{i \rightarrow a}) = 0.45\  \delta ( m_{i\rightarrow a}, 1)  + 0.45\  \delta( m_{i \rightarrow a}, -1)
  + 0.1 \ u( m_{i\rightarrow a} )$, where $u(x)$ is the uniform distribution
 over $-1 < x < 1$. This initial
  condition assumes that the spin value of a vertex $i$ is frozen in most of the macrostates.
\item[{\it U-type}.] The initial cavity magnetization distribution $P_{i\rightarrow a}(m_{i\rightarrow a})$ is set to be
$P_{i\rightarrow a} ( m_{i \rightarrow a}) = u( m_{i \rightarrow a} )$ with $-1 < m_{i \rightarrow a} < 1$. This condition assumes that initially the
spin value of a vertex is unfrozen in all the macrostates.
\end{description}
The plausibility of each of these two initial conditions will be judged by its predictions.

\section{Results}
\label{sec:RD}

\begin{figure}
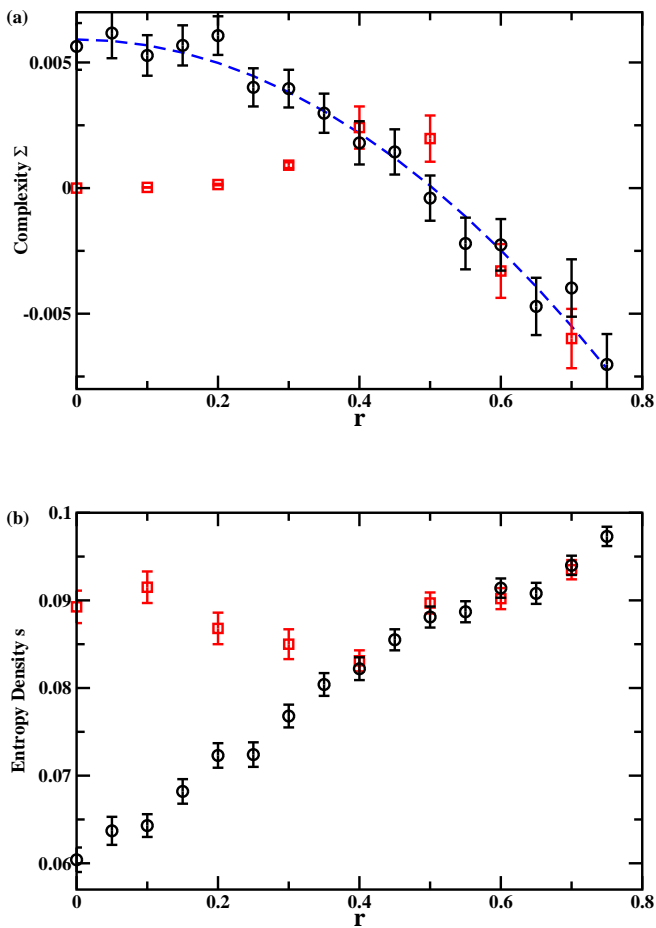

	\includegraphics[width=1.0\linewidth]{figure02a.eps}
	\vskip 1.0cm
	\includegraphics[width=1.0\linewidth]{figure02b.eps}
  \caption{\label{fig:Alpha4p2}
    (Color Online) The complexity (a) and mean entropy density (b) for the random $3$-SAT problem
    with constraint density $\alpha=4.2$.
    Black circles and red square are, respectively, simulation results obtained using
    the $F$-type and the $U$-type initial condition. The blue dashed line in (a) is a fitting to the
    circular points with $\Sigma(r) = 0.0059(2) - 0.023(1) \ r^2 $.
  }
\end{figure}

\subsection{Population dynamics at $\alpha=4.2$}

At $\alpha=4.2$, the complexity $\Sigma(r)$ and the mean entropy density 
$s(r)$ of the random $3$-SAT are shown in
Fig.~\ref{fig:Alpha4p2} for $0 \leq r \leq 0.75$. 
Under the $F$-type initial condition, the obtained complexity values can be fitted by
$\Sigma(r) = a - b  r^2$ with $a = 0.0059 \pm 0.0002$ and $b = 0.023 \pm 0.001$, while the
mean entropy density $s(r)$ increases monotonically from $s(0) = 0.060 \pm 0.001$ to
$s(0.75) = 0.097 \pm 0.001$.  These results appear to be quite reasonable: (i) According to
Refs.~\cite{Kirkpatrick-Thirumalai-PRB-1987,Monasson-1995,Zhou-Li-CTP-2008},
as $r$ increases, the complexity should decrease and the mean
entropy density should increase; (ii) the value of $\Sigma(0)$ agrees with the prediction of the
SP algorithm \cite{Mezard-Zecchina-2002}, which gives $\Sigma(0) = 0.00599$; (iii) $\Sigma(1)$ is
negative, in agreement with Ref.~\cite{Krzakala-etal-PNAS-2007}. The mean-field
theory suggests that the solution space of
a typical long random $3$-SAT formula with constraint density $\alpha = 4.2$ is dominated by
a few clusters of entropy density $s \approx  s(0.5) = 0.088 \pm 0.001$ [with $\Sigma(0.5) \approx 0$],
although 
clusters of lower entropy density $s=s(0) \approx 0.060$ are most abundant in the solution space. These two
entropy density values are in agreement with the results of Ref.~\cite{Montanari-etal-2008}.

When $r \geq 0.4$, the complexity and mean entropy density values reported by the population
dynamics with the $U$-type initial 
condition are in agreement with those obtained with the $F$-type initial 
condition.
For $r \geq 0.4$, the mean-field population dynamics is insensitive to initial conditions.
However, under the $U$-type initial condition the complexity $\Sigma(r)$ increases with $r$ and
the mean entropy density $s(r)$ decreases with $r$ when $r \in [0, 0.4)$. This behavior is
unphysical, because the mean entropy density $s(r)$ should be an increasing function
of $r$ \cite{Zhou-Li-CTP-2008}.
Therefore, under the $U$-type initial condition, the parameter $r$ should not be set to values 
lower than $0.4$. Under the $U$-type initial condition, the fixed point of the population dynamics at
$r=0$ corresponds to the replica-symmetric solution of the SP algorithm \cite{Mezard-Zecchina-2002}.
This replica-symmetric solution is always stable in the mean-field theory of
Ref.~\cite{Mezard-Zecchina-2002}, as entropic effects are completely neglected.
When the entropy of each zero-energy macrostate is properly considered in the mean-field
theory, the present paper indicates that this replica-symmetric solution is no longer stable
(see also Refs.~\cite{Krzakala-etal-PNAS-2007,Montanari-etal-2008}).
To get physically meaningful results for $0 \leq r < 0.4$
under the $U$-type initial condition, it is necessary to assume further organization of the 
solution space of the random $3$-SAT problem (splitting of a cluster of solutions into
many subclusters of solutions). Implementing this higher-order hierarchical structure into
the population dynamics is conceptually simple, but the algorithm  will 
be extremely demanding on computer time and memory space.

\subsection{Population dynamics at $\alpha = 4.2667$}

\begin{figure}
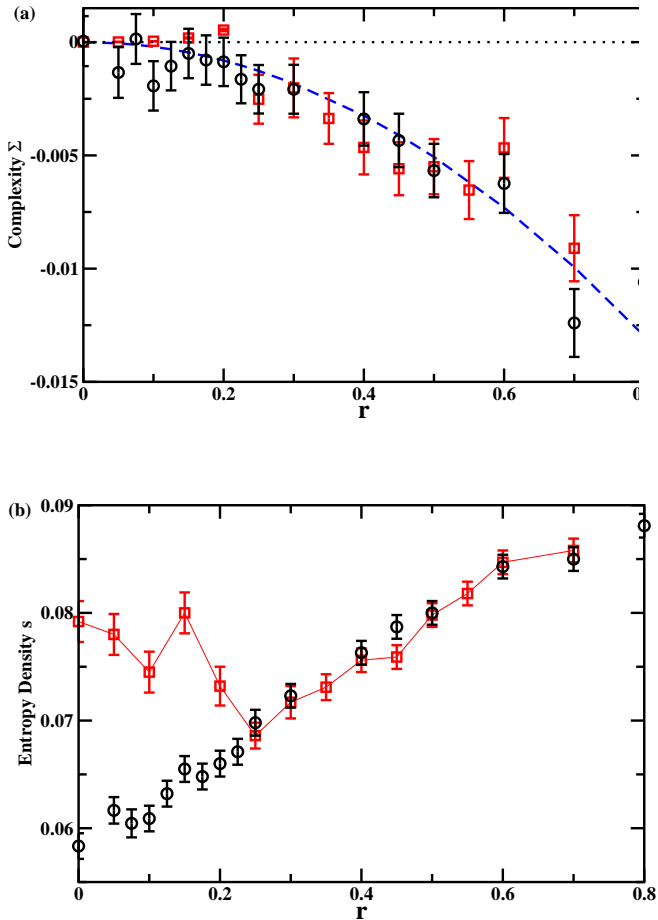

	\includegraphics[width=1.0\linewidth]{figure03a.eps}
	\vskip 1.0cm
	\includegraphics[width=1.0\linewidth]{figure03b.eps}
\caption{\label{fig:Alpha4p2667}
  (Color Online) The complexity (a) and mean entropy density (b) for the random $3$-SAT
  problem with constraint density $\alpha=4.2667$. 
  Black circles and red square are, respectively, simulation results obtained using
  the $F$-type and the $U$-type initial condition. The blue dashed line in (a) is a fitting to the
  circular points with $\Sigma(r) = - 0.020(1) \ r^2 $, while the black dotted line marks
  $\Sigma(r)\equiv 0$.}
\end{figure}

The SAT-UNSAT transition of the random $3$-SAT problem is predicted to occur at
$\alpha = 4.2667$ \cite{Mezard-Zecchina-2002,Mertens-etal-2006}. At this density of
constraints, Fig.~\ref{fig:Alpha4p2667} shows how the complexity and entropy
density change with the ratio $r$. Under the $F$-type initial condition, 
the complexity decreases with $r$ according to $\Sigma(r) = - b  r^2$ with $b = 0.020 \pm 0.001$;
and consistently, the entropy density $s(r)$ increases with $r$ monotonically. 
The present work, therefore, further confirms that the satisfiability transition of the
random $3$-SAT takes place at $\alpha = 4.2667$: including the entropic effect into the mean-field
theory does not change the predicted location of the SAT-UNSAT transition. 
At this transition point, a typical random
$3$-SAT formula of length $N$ still has an exponential number $\exp\bigl[N s(0) \bigr]$ of
satisfiable solutions, with $s(0) = 0.058 \pm 0.001$. But it is extremely difficult for
a local or global algorithm to find one such solution.

As in the case of $\alpha=4.2$, if the $U$-type initial condition is applied,
both the calculated complexity $\Sigma(r)$ and mean entropy density $s(r)$ do not change monotonically
with $r$. Figure~\ref{fig:Alpha4p2667} indicates that the results from the $U$-type initial condition
are valid only for $r \geq 0.25$. For $0 \leq r \leq 0.2$, as the increasing trend
of the complexity $\Sigma(r)$
and the decreasing trend of the entropy density $s(r)$ are not physically meaningful, the positivity
of $\Sigma(r)$ cannot be taken as evidence that the random $3$-SAT is still in the SAT phase at
$\alpha=4.2667$.

\subsection{Population dynamics at $\alpha=4.0$}

\begin{figure}[t]
	\includegraphics[width=1.0\linewidth]{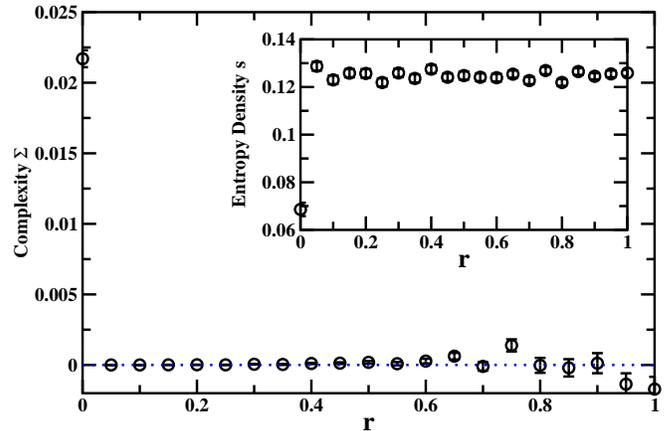}
\caption{\label{fig:Alpha4p0}
  (Color Online) The complexity and mean entropy density (inset) for the random $3$-SAT
  problem with constraint density $\alpha=4.0$.  Black circles are simulation results obtained using
  the $F$-type initial condition. The blue dashed line marks $\Sigma(r) \equiv 0$.}
\end{figure}

For $\alpha=4.2667$ and $4.2$, the complexity $\Sigma(r)$ calculated with
the $F$-type initial condition reaches maximum at $r=0$ and it 
has the form $\Sigma(r) = a - b  r^2$ when $r \in [0, 1)$.
However, Fig.~\ref{fig:Alpha4p0} demonstrates that a different situation
occurs for $\alpha = 4.0$. At this density of constraints, the population dynamics with $r=0$ 
and the $F$-type initial condition reports a complexity value $\Sigma(0) = 0.0217 \pm 0.0006$ (agreeing
with the prediction of the SP algorithm \cite{Mezard-Zecchina-2002}) and a
mean entropy density value $s(0) = 0.069 \pm 0.003$.
But as the ratio $r$ is set to slightly positive values,
the complexity suddenly drops to $\Sigma(r) \approx 0$ while the mean entropy density jumps to $s(r) \approx 0.125$.
As $r$ increases further, both $\Sigma(r)$ and $s(r)$ keep almost constant until $r$ is close to unity.
For $r \geq 0.8$, $\Sigma(r)$ and $s(r)$ have, respectively, a decreasing and an increasing trend.
The discontinuity at $r=0$ for both $\Sigma(r)$ and $s(r)$ was totally unexpected 
(we have performed population dynamics simulations with different $F$-type 
initial conditions to rule out
the possibility of numerical artifact). Similar discontinuity was also observed in the
$q$-coloring problem \cite{Zdeborova-Krzakala-PRE-2007}.
If we look at the steady-state cavity magnetization distributions $P_{i\rightarrow a}(m_{i\rightarrow a})$, we find that they
are far from being in the form of a $\delta$-function in the whole range of $0 \leq r \leq 1$. This later
observation confirms that at $\alpha = 4.0$, the ergodicity property of the solution space of the
random $3$-SAT is indeed violated. 
Figure~\ref{fig:Alpha4p0} indicates that at $\alpha = 4.0$, the solution space
of the random $3$-SAT problem is organized far more complex than what has been assumed in the
mean-field theory. This point should be investigated more thoroughly.

For the limiting case of $r=0$, it has already been shown that the mean-field solution at the first-step
replica-symmetry-broken level is unstable toward the full-step replica-symmetry-broken level
\cite{Montanari-etal-2004,Mertens-etal-2006} for $\alpha < 4.153$. The different behaviors demonstrated
in Figs.~\ref{fig:Alpha4p2}, \ref{fig:Alpha4p2667}, and \ref{fig:Alpha4p0} for
$\alpha = 4.2, 4.2667$, and $4.0$ confirm the earlier stability analysis
\cite{Montanari-etal-2004,Mertens-etal-2006} and further suggest that, if the 1RSB mean-field solution
is unstable at $r=0$, it will be unstable when $r$ is positive but less than a certain threshold
value $r_{\rm th}$. This threshold value may be smaller or larger than unity (for $\alpha = 4.0$, it appears that
$r_{\rm th} \approx 0.8$).

\subsection{Population dynamics at $r=1$}

\begin{figure}
\includegraphics[width=1.0\linewidth]{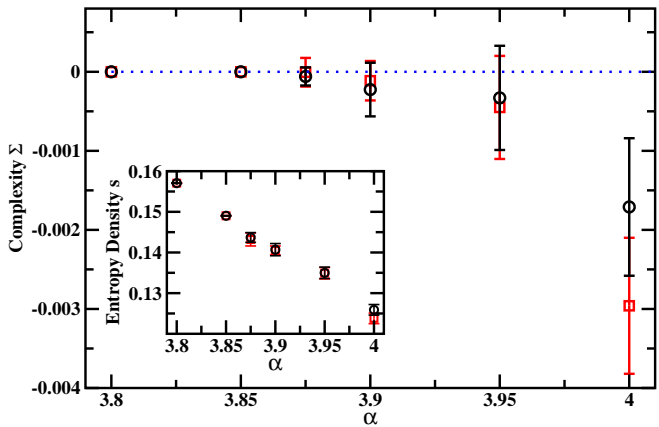}
\caption{\label{fig:R1}
  The complexity $\Sigma(r=1)$ and the mean entropy density $s(r=1)$  for the random $3$-SAT problem with
  constraint density $\alpha$.  Black circles and red squares correspond to the $F$-type and the 
  $U$-type initial
  condition, respectively. The blue dotted line marks $\Sigma \equiv 0$.}
\end{figure}

Now let us fix $r=1$ and study how the complexity $\Sigma(1)$ and mean entropy density 
$s(1)$ change with the constraint density $\alpha$. Using an elegant tree reconstruction technique,
Montanari and co-authors \cite{Krzakala-etal-PNAS-2007} found that, for the
random $3$-SAT problem, $\Sigma(1)$ changes from being exactly zero to being negative at $\alpha \approx 3.87$. 
The alternative population dynamics approach of the present paper reports consistent results
(see Fig.~\ref{fig:R1}).  For $\alpha = 3.8$ and $3.85$, we have checked that the steady-state
distributions $P_{i\rightarrow a}(m_{i\rightarrow a})$ of cavity magnetizations are all $\delta$-functions (ergodicity property
of the solution space is not violated). For $\alpha \geq 3.875$, simulations with both the $F$-type and
$U$-type initial
condition give negative values for the complexity $\Sigma(1)$. At $\alpha$ very close to the ergodicity
transition point of $3.87$, we have also observed that the population dynamics simulation needs a much longer
time to reach steady state.  This behaviors is very probably caused by the divergence of
relaxation times of the population dynamics at the vicinity of the ergodicity transition ($\alpha
\approx 3.87$). Such a critical slowing-down was investigated analytically and numerically
in Ref.~\cite{Weigt-Zhou-2006}.

When $\alpha > 3.87$, very probably most of the satisfying solutions of a 
random $3$-SAT formula can be grouped into one of
a subexponential number of clusters of solutions \cite{Krzakala-etal-PNAS-2007,Zhou-Li-CTP-2008}.
It will then be very difficult to prove mathematically the clustering of solutions following
the method of Ref.~\cite{Mezard-etal-2005}.

\section{Conclusion and discussion}
\label{sec:CO}

In this paper, we studied a spin glass model of the random $3$-SAT problem
at the temperature $T\rightarrow 0$ limit by the mean-field first-step replica-symmetry-breaking
(1RSB) cavity method. The reweighting parameter $y$ (corresponding to the level of macrostates) 
and the inverse temperature $\beta$  were
allowed to approach infinity with fixed ratio $r = y / \beta$.
The complexity and mean entropy density
of the random $3$-SAT are calculated as a function of $r$ by population dynamics simulations.
The sensitivity to initial conditions of the simulation results was investigated by initializing
the cavity magnetization distributions in two different way (see Sec.~\ref{sec:populationdynamics}).

When the $F$-type initial condition  is used, at $\alpha = 4.2$ the complexity $\Sigma(r)$ decreases
monotonically with $r$ and becomes negative when $r$ exceeds $0.5$; the mean entropy density
$s(r)$ increases monotonically with $r$. The most abundant clusters of solutions of the random 
$3$-SAT system correspond to $r=0$ and have mean entropy density $s(0)\approx 0.060$,
but the (few) dominating clusters of solutions correspond to $r \approx 0.5$ and have mean entropy
density $s(0.5) \approx 0.088$.  The complexity $\Sigma(r=0)$ decreases continuously with $\alpha$ and
reaches zero at $\alpha=4.2667$, where the random $3$-SAT experiences a SAT-UNSAT
transition. At this critical constraint density, the solution space of the random $3$-SAT
still has a positive mean entropy density $s(0) \approx 0.058$.

When the $U$-type initial condition is applied, the complexity $\Sigma(r)$ and mean entropy
density $s(r)$ are both  nonmonotonic functions of $r$. At $\alpha = 4.2$, the population dynamics
algorithm reported a zero complexity value at $r=0$. As $r$ becomes positive, 
$\Sigma(r)$ first increases with $r$, 
reaches a maximal value at $r\approx 0.4$, and then decreases with $r$. The mean entropy density $s(r)$
has a reverse trend. The non-monotonic behaviors of $\Sigma(r)$ and $s(r)$ indicate that, for
the $U$-type initial condition the population dynamics will not report physically meaningful
results if $r$ is close to zero. At $\alpha = 4.0$, if the parameter $r$ is set close to zero,
even the population dynamics with the $F$-type
initial condition will fail to get plausible results.

At $r = 1$, the complexity and mean entropy density as a function of constraint density 
$\alpha$ were also investigated by population dynamics. For $\alpha = 3.85$ or lower, ergodicity of the
solution space of the random $3$-SAT is unbroken and the complexity is exactly zero.  For
$\alpha =3.875$ or higher, the population dynamics with both the $F$-type and the $U$-type initial
condition predicted negative values for $\Sigma(1)$. The zero-energy configuration space of the
random $3$-SAT problem clusters into many subspaces for $\alpha > 3.875$, but only 
subexponential clusters are dominating the configuration space,
in agreement with Ref.~\cite{Krzakala-etal-PNAS-2007}.

This paper focused on the zero-energy configurational space of the random $3$-SAT problem.
When the ground-state energy of the system becomes positive, the $T\rightarrow 0$ limit
formulas in Sec.~\ref{sec:populationdynamics} need to be revised. Most importantly, in a given
macrostate a cavity magnetization $m_{i\rightarrow a}$ may  take one of the following three possible forms:
\begin{equation}
  \label{eq:newexpression}
m_{i \rightarrow a} = \left\{
\begin{array}{l}
1- m_{i \rightarrow a}^+ e^{-2 \beta} \\
\; \\
-1 + m_{i \rightarrow a}^{-} e^{-2 \beta} \\
\; \\
m_{i \rightarrow a}^{0}
\end{array}
\right.
\end{equation}
where $-1 < m_{i\rightarrow a}^{0} < 1$, $m_{i\rightarrow a}^{+} \geq 0$, and $m_{i\rightarrow a}^{-} \geq 0$. 
In the present paper, we have
simply set $m_{i \rightarrow a}^{+} = m_{i\rightarrow a}^{-} = 0$ 
without affecting the results of population dynamics, but
for systems with positive ground-state energies, the more general formula should be
used. Even if the ground-state energy of the system is zero, Eq.~(\ref{eq:newexpression}) should be used if
one wants to study the properties of metastable macrostates (with positive minimal energies) or the low-temperature
properties of the system. We will return to this point in a later publication. 

As Refs.~\cite{Krzakala-etal-PNAS-2007,Montanari-etal-2008} and the present paper demonstrate,
the zero-energy configuration space of the random $3$-SAT problem is divided into clusters of different sizes.
For the random $3$-SAT problem, will the minimal-energy configurations with a given positive energy
value $E$ also be split into clusters of different entropies $S$? To detect such a possibility, a
natural extension is to introduce two reweighting parameters (say $y$ and $r$) for both energy and
entropy, and to reweight each minimal-energy cluster $\alpha$
by a factor $\exp(- y E + r S)$. Together with Krzakala and Zdeborova, we are
working on this point for the random $3$-SAT problem and the $q$-coloring problem.

Although physicists believe that the solutions  of a large random $3$-SAT formula are
organized into well separated subspaces,
clustering of random $K$-SAT solutions has been rigorously proven only
for $K \geq 8$ \cite{Mezard-etal-2005-a}.
Recently, there has been a lot of simulation work on this important issue
(e.g., \cite{Ardelius-etal-JSM-2007,Ardelius-Zdeborova-2008}),
but a lot of work still remains to be done to fully understand
the energy landscape of the random $3$-SAT problem.

\section*{Acknowledgment}

The author thanks Pan Zhang for computer resources. The hospitality of Tie-Zheng Qian
(Mathematics Department, Hong Kong University of Science and Technology) is gratefully
appreciated. The author also thanks Erik Aurell,
Florent Krzakala, and
Lenka Zdeborova for helpful discussions.
This work is partially supported by NSFC (Grant No. 10774150).

\end{document}